\documentclass[12pt,showpacs,preprint,nofootinbib]{revtex4}
\usepackage[intlimits]{amsmath}
\usepackage{amssymb}
\usepackage{exscale}
\usepackage{graphicx}
\usepackage{setspace}

\newcommand{\ud}{\mbox{d}}
\newcommand{\CL}{\mathcal{L}}
\newcommand{\CD}{\mathcal{D}}
\newcommand{\CE}{\mathcal{E}}
\newcommand{\h}{\frac{1}{2}}
\newcommand{\matlab}{\textsc{Matlab}\textsuperscript{\textregistered}}

\begin{document}

~

\vspace{2 cm}

\title {Spinning Q-Balls}

\preprint{FSU TPI 01/02}
\preprint{hep-th/0205157}

\author{Mikhail S. Volkov\footnotemark[2]}
\author{Erik W\"ohnert\footnotemark[3]}

\vspace{1 cm}

\affiliation{\footnotemark[2]Laboratoire de Math\'ematiques et Physique
Th\'eorique, Universit\'e de Tours, Parc de Grandmont, 37200 Tours,
FRANCE \footnotetext[2]{\tt volkov@phys.uni-tours.fr} \\
\footnotemark[3]Theoretisch-Physikalisches Institut, Friedrich
Schiller Universit\"at Jena, Fr\"obelstieg 1, 07743 Jena,
GERMANY \footnotetext[3]{\tt pew@tpi.uni-jena.de}}

\vspace{1 cm}

\begin{abstract}

\noindent
We present numerical evidence for the existence of spinning
generalizations for 
non-topological Q-ball solitons in the theory of a complex scalar field
with a non-renormalizable self-interaction.
To the best of our knowledge, this provides the first explicit example
of spinning solitons in $3+1$ dimensional Minkowski space.
In addition, we find an infinite discrete family of radial excitations of
non-rotating Q-balls,
and construct also spinning Q-balls in $2+1$ dimensions.
\end{abstract}

\pacs{11.27.+d}
\keywords{Q-balls, solitons}
\maketitle

\section{Introduction}

Solitons are important ingredients of models in high energy physics.
Apart from 
being responsible for various non-perturbative quantum phenomena,
solitons are interesting in themselves, since they can be viewed as
field theoretic realizations 
of elementary particles. It is from this viewpoint that
solitons were originally introduced into physics in the context of the
Skyrme model
-- as models of hadrons. The subsequent developments have revealed
soliton solutions in many other non-linear field theories in Minkowski
space, such as 
monopoles, vortices, sphalerons, Q-balls, etc. These solutions describe
localized, particle-like objects with finite energy. Their spectra of
energy and charge are typically discrete. In addition, these solutions
are regular everywhere, a property which is especially appealing. So
far, however, solitons have been lacking one important feature of
elementary particles: the intrinsic angular momentum $J$, which is zero
for all known classical solutions.

It is sensible to ask whether \textit{stationary} rotating
generalizations for the known static soliton solutions exist. More
precisely, 
one is interested in finite energy, globally regular, non-radiating
solutions for which the spatial integral of the $T_{0\varphi}$ component
of the energy-momentum tensor,
\begin{eqnarray}
\label{J}
J & = & \int T_{0\varphi} \, \ud^3 x \;,
\end{eqnarray}
is non-vanishing. In this definition the condition of stationarity
(absence of radiation) is important. Indeed, it is always possible to
construct a field configuration such that $J \neq 0$ at the initial
moment of time. Physically this would correspond to exciting the soliton
to give it an angular momentum%
\footnote{In the literature one can often find explicit examples of
`solitons in the rigid rotator approximation',
for which the integral (\ref{J}) does
not vanish, as for instance `rotating' Skyrmions
\cite{Adkins:1983ya}, knots \cite{Gladikowski:1997mb}, etc.  These
configurations, however, are not solutions of the equations of motion,
and at best they can be viewed 
as the initial values for the dynamical evolution problem.}.
However, when the time evolution starts, the received excitation will be
most probably immediately radiated away, leaving behind a non-rotating
object.

When talking about rotating solitons, it also seems sensible to
distinguish between two types of rotation: \textit{spinning} and
\textit{orbiting}. \textit{Spinning} is associated with the
intrinsic angular momentum, in analogy with the quantum-mechanical spin.
Classical spin excitations, if they exist, should live in the
one-soliton sector because they are excitations of an individual
object. One expects that the corresponding angular momentum will assume
only discrete values. Solutions describing spinning solitons in
Minkowski space in $3+1$ dimensions are not known. For example, it is
not known whether the 't Hooft-Polyakov monopoles can be given classical
spin\footnote{There is also the possibility to associate the spin
of the monopole with that of fermionic zero modes living in the
monopole background. In this case, however, the spin is not classical,
and in fact is not related to the monopole itself.}. 
In fact, it has been shown that, at the \textit{perturbative}
level, monopoles do not admit stationary rotational excitations
in the one-monopole sector
\cite{Heusler:1998ec}. This means that monopoles
cannot rotate \textit{slowly}, with $|J| \ll 1$. Monopoles with finite
(discrete) values of $J$ are not yet excluded. 
However, to decide
whether such solutions exist requires to go beyond perturbation theory
and solving the complete coupled system of the Yang-Mills-Higgs (YMH)
partial differential equations (PDEs), which is an exceedingly
difficult task%
\footnote{It was argued in \cite{VanderBij:2001nm} 
that at least within the minimal axial ansatz,
rotating 't Hooft-Polyakov monopoles can be excluded also
at the non-perturbative level.}.

On the other hand, one can consider relative \textit{orbital} motions in
composite many-soliton systems.  For example, one can imagine a rotating
soliton-antisoliton pair balanced against mutual attraction by the
centrifugal force. The spectrum of the angular momentum is then expected
to be continuous. Solutions describing such orbiting solitons can
actually be constructed. In the case of monopoles, for instance, there
exist \cite{Taubes:1982ie,Taubes:1982if} solitonic solutions of the YMH
field equations with vanishing Higgs potential in the sector with zero
monopole charge. Such static, purely magnetic solutions have been
explicitly constructed in \cite{Kleihaus:1999sx}; they describe
monopole-antimonopole pairs balanced by a repulsive force of
topological nature. Now, there is a simple way to add angular
momentum to these solutions \cite{Heusler:1998ec} by using the global
symmetry of the field equations which mixes the Higgs field $H$ and the
electric potential $\Phi$:
\begin{eqnarray}
\label{hyp}
H & \to & H \cosh \gamma + \Phi \sinh \gamma \;, \qquad 
\Phi \; \to \; \Phi \cosh \gamma + H \sinh \gamma \;.
\end{eqnarray}
Applying this transformation with an arbitrary $\gamma \neq 0$ leads to
solutions with an electric field. It is important to note that this also
changes the angular momentum to 
$J \sim \sinh \gamma$ \cite{Heusler:1998ec}%
\footnote{This trick works only for non-BPS solutions -- those obeying
the second order YMH equations but not the first order
Bogomol'nyi equations. For BPS solutions the angular momentum is
invariant under (\ref{hyp}).  For this reason one cannot use this method
to produce spinning monopoles, because non-BPS solutions with 
finite energy and unit topological charge are not known
\cite{Maison:1981ze}.} 
(in fact  
$J=Q$, where $Q$ is the electric charge \cite{VanderBij:2001nm}). 
These new solutions can be interpreted as describing the system of a
monopole and antimonopole rotating around their common center of mass.

Another example of systems which could be classified as orbiting
solitons are rotating vortex loops. In certain models there exist vortices 
with stationary currents along them; `superconducting
vortices'. One can argue \cite{Davis:1988ip,Davis:1989ij} that taking a
finite piece of such a vortex, bending and closing it to form a loop,
finally adding a momentum along the loop, leads to an object (vorton)
described by a stationary solution of the equations of
motion\footnote{To our knowledge such solutions have not been
constructed explicitly.}. The angular momentum in this case is
associated with the macroscopic circulation of the current along the
loop. This type of motion could be naturally classified as 
orbital rotation.

On the other hand, one can also consider the rotation of a straight
vortex \textit{along} its symmetry axis. This would 
correspond more closely to the notion of an intrinsic spinning
rotation. Although one can show that for the Nielsen-Olesen vortex such
spinning excitations do not exist, they can actually exist in other
models
\cite{deVega:1986eu,Jackiw:1990aw,Kim:1993mm,Piette:1995mh,Gisiger:1997vb}. 
However, this only 
gives spinning solitons in $2+1$ dimensions, while their $3+1$
dimensional analogs will have infinite energy due to the infinite length
of the vortex.

In summary, we are not aware of any spinning solitons in Minkowski
space in $3+1$ dimensions. At the same time, such solutions are known in
curved space. Indeed, there are many rotating solutions in General
Relativity. In the pure gravity case they comprise the family of
Kerr-Newman black holes. These are very similar to solitons, but
they are not globally regular and contain a curvature
singularity. There are also globally regular rotating solutions
\cite{Neugebauer}, but these require a source of a non-field theoretic
origin. Interestingly, there exist
two explicit examples of gravitating spinning solitons in pure field
systems. 

The first example is provided by rotating boson stars 
\cite{Schunk96,Yoshida:1997qf}.
These are solutions for a gravity-coupled massive complex scalar field
with harmonic dependence on time and on the azimuthal angle,
$\Phi(t,r,\vartheta,\varphi) = \exp(i \omega t + i N \varphi)
f(r,\vartheta)$, with $N$ integer. The energy-momentum tensor is
time-independent, and the Einstein equations together with the
Klein-Gordon equation for $f(r,\vartheta)$ admit globally regular,
stationary particle-like solutions \cite{Schunk96}. The angular momentum
is quantized as $J \sim N$. The solutions with $N > 0$ can be regarded
as spinning excitations of the fundamental static, spherically symmetric
solutions with $N = 0$.

This example is instructive in the sense that it is clear
`what rotates': this is the phase $\omega t + N \varphi$. In
this connection it is worth noting that the very notion of rotation in
pure field systems is very different from that for ordinary rigid
bodies. Indeed, it is meaningless to say that a given element of volume
of a field system actually `performs revolutions' around a given axis.
Of course, one can imagine a solitonic object with a field perturbation
running around it. However, such a moving perturbation will be most
probably immediately radiated away. The example of boson stars thus
shows that one can nevertheless have a rotating phase which is not
radiated away.

In stationary rotating systems without explicit time dependence the
rotation will rather be associated with certain non-linear
superpositions of the multipole moments of the fields.  For example, in
systems with vector fields, angular momentum may be present
due to a non-vanishing integral involving the Poynting vector,
\begin{eqnarray}
\label{1}
\vec{J} & = & \int \vec{r}\times(\vec{E}\times\vec{B}) \, \ud^3 x \;. 
\end{eqnarray} 
A stationary, globally regular configuration for which this
integral is non-zero would correspond to a rotating
soliton.

It is possible that such solutions could exist for the 
SU(2) Yang-Mills fields coupled to gravity. The Einstein-Yang-Mills
(EYM) field equations admit globally regular, particle like solutions
\cite{Bartnik:1988am}. These gravitating EYM particles are static,
spherically symmetric and neutral (their purely magnetic gauge field
strength decays asymptotically as $1/r^3$). Now,  
for these solutions one can {\it perturbatively} construct stationary, 
globally regular,
axially-symmetric, slowly rotating generalizations 
\cite{Brodbeck:1997ek}. Surprisingly, their spectrum of $J$ is
\textit{continuous}, as if they were composite objects, which is 
presumably due to the special feature of the
field system consisting of only \textit{massless} physical fields.
Unfortunately, 
it is not clear at the moment whether these perturbative 
solutions exist also at the 
non-perturbative level -- the analysis of 
\cite{VanderBij:2001nm,Kleihaus:2002ee} indicates
the opposite. 
It is however still possible that 
the solitons exist, but  
within a more general ansatz 
than that considered in \cite{VanderBij:2001nm,Kleihaus:2002ee}.

In summary, spinning solitons have been
found only in curved space. It is then sensible to ask if spinning solitons
without gravity exist at all. In principle, it is not excluded that only
gravity can support the relevant rotational degrees of freedom. In order
to rule out this logical possibility, we have undertaken an attempt to
construct spinning solitons in 
flat space.

The solitons we have chosen to `rotate' are Q-balls
\cite{Coleman:1985ki,Lee:1992ax}. These are solutions for a complex
scalar field with a non-renormalizable self-interaction arising
in some effective field theories. These solutions circumvent the
standard Derrick-type argument due to having a time-dependent phase for
the scalar field. In this sense they are somewhat similar to the boson
stars. In the simplest spherically symmetric case the fundamental Q-ball
solution was described by Coleman \cite{Coleman:1985ki}. Dynamical
properties of these objects have been studied in \cite{Battye:2000qj}.
Q-balls also appear in supersymmetric generalizations of the standard
model, where one finds leptonic and baryonic Q-balls
\cite{Kusenko:1997zq} which may be responsible for the generation of
baryon number or may be regarded as candidates for dark
matter \cite{Kusenko:1998si}.

In the present paper, we first recall 
the properties of the fundamental spherically symmetric Q-balls. In
addition, we present an infinite family of their radial excitations
which have not been reported in the literature before. We then turn to
rotating solutions and first consider them in $2+1$ spacetime
dimensions, where the problem reduces to an ordinary differential
equation. Finally we consider the full problem in $3+1$ dimensions and
explicitly construct solutions with angular momentum. To our knowledge,
our analysis gives the first explicit example of spinning solitons in
flat space.

\section{The Model}

\begin{figure}[ht]
\begin{minipage}[b]{0.45\linewidth}
  \centering
  \includegraphics[width=\linewidth]{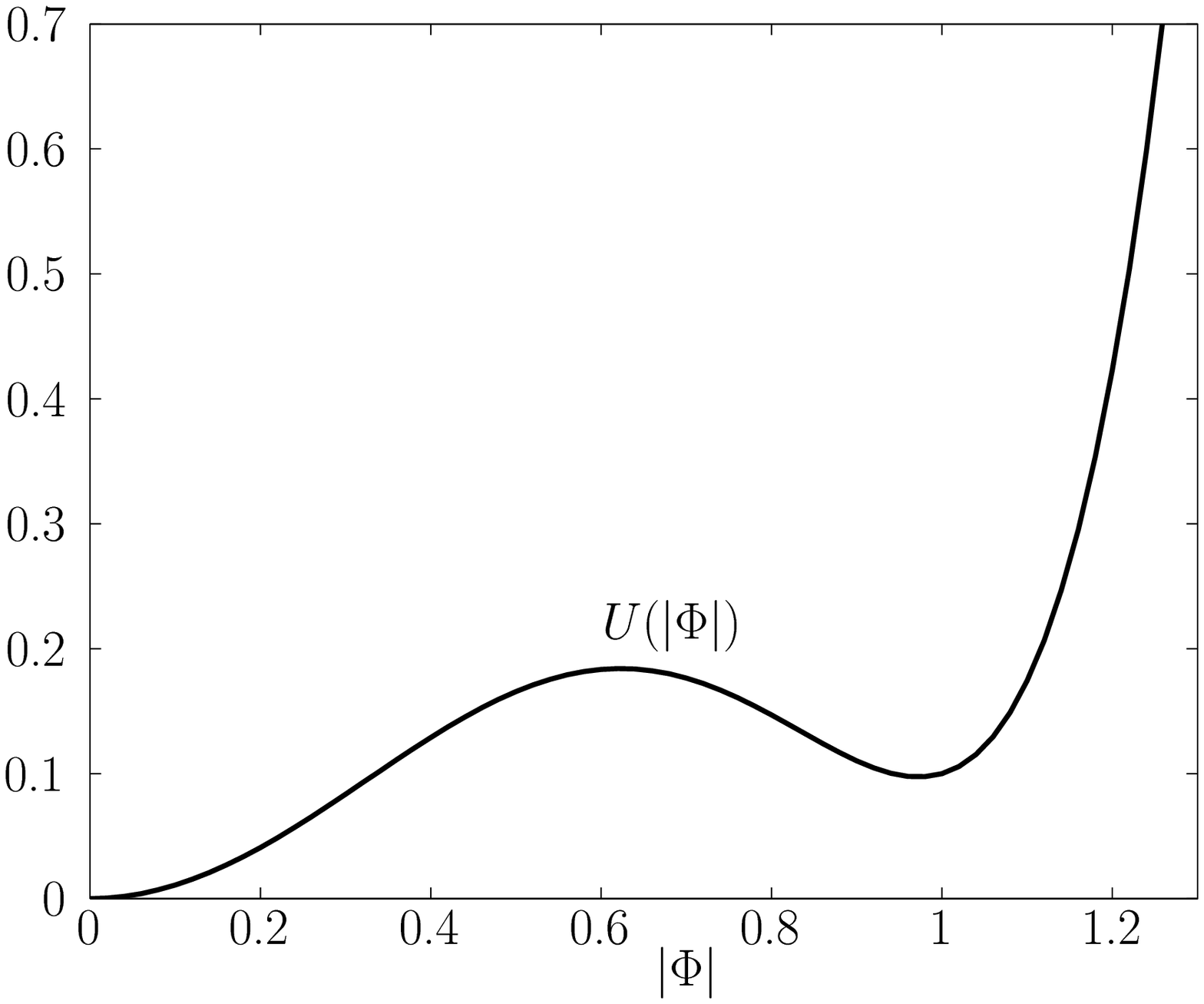}
  \caption{ The qualitative shape of the potential $U(|\Phi|)$.}
  \label{fig1}
\end{minipage}
\hspace{5 mm}
\begin{minipage}[b]{0.45\linewidth}
  \centering
  \includegraphics[width=\linewidth]{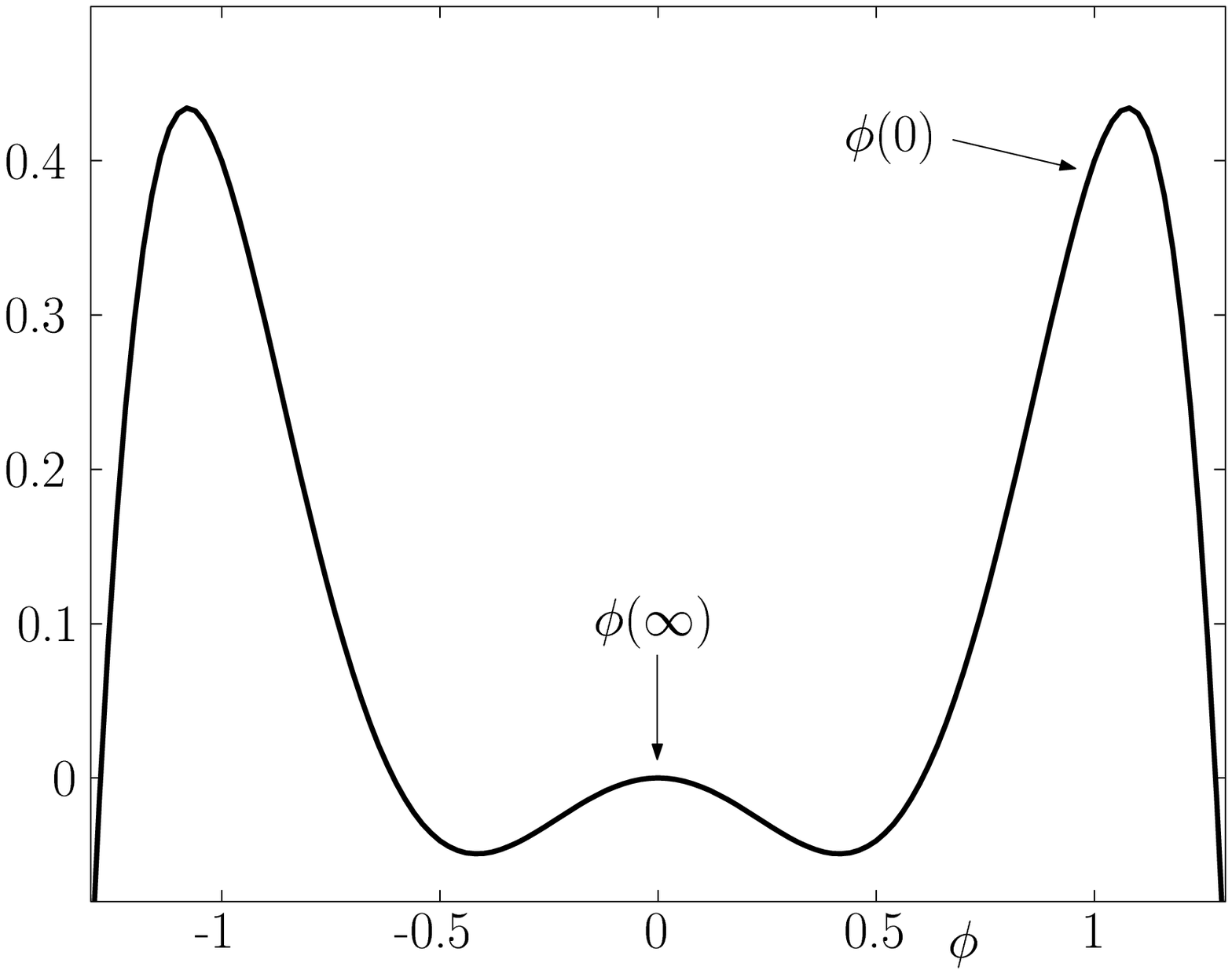}
  \caption{The effective potential   $V(\phi) = \frac{1} {2} \omega^2
  \phi^2 - U(\phi)$ in Eq.~(\ref{radeom2}).}
  \label{fig2}
\end{minipage}
\end{figure}

Let us consider a theory of a complex scalar field in $3+1$ dimensions
defined by the Lagrangian
\begin{eqnarray}
\label{lagdens}
{\CL} & = & \partial_{\mu} \Phi \, \partial^{\mu} \Phi^\ast - U(|\Phi|)
\;.
\end{eqnarray}
It is assumed that the potential $U$ has its global minimum at $\Phi =
0$, where $U(0) = 0$, while $U \to \infty$ for $|\Phi| \to \infty$. In
addition,  the potential must fulfill a particular
inequality (Eq.(\ref{cond3})) which will be discussed below. 
The potential may also have local minima at some finite $|\Phi|$, as is
shown in Fig.~\ref{fig1}, but this is not necessary.

The global symmetry of the Lagrangian under $\Phi \rightarrow \Phi
e^{i \alpha}$  gives rise to the conserved charge 
\begin{eqnarray}
\label{charge1}
Q & = & \frac{1}{i} \int \ud^3 x \, (\Phi^{*} \dot{\Phi} - \Phi
\dot{\Phi}^{*}) \;.
\end{eqnarray}
The fundamental Q-ball solutions of the theory are minima of
the energy for a given $Q$ \cite{Coleman:1985ki}. Since $\Phi$ should
depend on time for $Q$ to be non-vanishing, one assumes that $\Phi$
has a harmonic time dependence. In the spherically symmetric case,
\begin{eqnarray}
\label{radphi}
\Phi & = & \phi(r) \, e^{i \omega t} \;, 
\end{eqnarray}
where $\phi(r)$ is real. The potential $U(|\Phi|) = U(\phi)$ and the
energy-momentum tensor,
\begin{eqnarray}
\label{radt}
T_{\mu\nu} & = & \partial_\mu \Phi \, \partial_\nu \Phi^\ast +
\partial_\nu \Phi \, \partial_\mu \Phi^\ast - g_{\mu\nu} \, \CL \;,
\end{eqnarray}
($g_{\mu\nu}$ being the spacetime metric) do not depend on time. The
energy distribution is therefore stationary, and the total energy is
\begin{eqnarray}
\label{radenergy}
E & = & 4 \pi \int_0^\infty \ud r \, r^2 (\omega^2 \phi^2 +
\phi^{\prime 2} + U(\phi)) \;,
\end{eqnarray}
where the prime denotes differentiation with respect to $r$. The field
equation,
\begin{eqnarray}
\label{eq}
0 & = & \frac{1} {\sqrt{-g}} \partial_\mu (\sqrt{-g} \, g^{\mu\nu}
\partial_\nu \Phi) + \frac{\partial U} {\partial \Phi^\ast} \;,
\end{eqnarray}
reduces to
\begin{eqnarray}
\label{radeom}
0 & = & \phi'' + \frac{2} {r} \, \phi' - \frac{\ud U(\phi)} {\ud\phi} +
\omega^2 \phi \;,
\end{eqnarray}
which is equivalent to 
\begin{eqnarray}
\label{radeom2}
\h \, \phi^{\prime 2} + \h \,\omega^2 \phi^2 - U & = & \CE - 2 \int_0^r
\frac{\ud r} {r} \, \phi^{\prime 2} \;. 
\end{eqnarray}
This effectively describes a particle moving with friction in the one
dimensional potential 
\begin{eqnarray}
V(\phi) = \h \, \omega^2 \phi^2-U(\phi) \;.
\end{eqnarray}
$\CE$ is an integration constant playing the role of the total
`effective energy'. It is essential that the potential $V(\phi)$ should
have the qualitative shape shown in Fig.~\ref{fig2}, which is possible if
the following conditions are fulfilled. First, since $V''(0) < 0$, it
follows that $\omega^2$ should not be too large:
\begin{eqnarray}
\label{cond1}
\omega^2 & < & \omega^{2}_{\max} \; \equiv \; U''(0) \;. 
\end{eqnarray}
On the other hand, $\omega^2$ should not be too small, since otherwise
$V(\phi)$ will be always negative. $V(\phi)$ will become positive for
some non-zero $\phi$, as is shown in Fig.~\ref{fig2}, if only 
\begin{eqnarray}
\label{cond2}
\omega^2 & > & \omega^{2}_{\min} \; \equiv \; \min_{\phi}
({2U(\phi)}/{\phi^2})
\;,
\end{eqnarray}
where the minimum is taken over all values of $\phi$. For the potential
$U(\phi)$ it is necessary that
\begin{eqnarray}
\label{cond3}
U''(0) & > & \min_{\phi} ({2U(\phi)}/{\phi^2}) \;,
\end{eqnarray}
since only then the set of values of $\omega^2$ will be non-empty. The
only possible renormalizable interaction in the theory, $U = \h \mu^2
|\Phi|^2 + \lambda |\Phi|^4$, does not obey this condition.
Thus non-renormalizable potentials have to be considered
\cite{Coleman:1985ki}. For example, for the potential
\begin{eqnarray}
\label{polynpot}
U(\phi) & = & \lambda (\phi^6 - a \phi^4 + b \phi^2)
\end{eqnarray}
the condition (\ref{cond3}) is fulfilled for any positive $\lambda$,
$a$, $b$, and $U(\phi)$ will have a global minimum at $\phi = 0$ if $b >
a^2/4$. We use this model potential with $\lambda=1$, $a=2$ and $b=1.1$
in all our calculations below, in which case the conditions
(\ref{cond1}),~(\ref{cond2}) require that $0.2 \leq \omega^2 \leq 2.2$.

\section{Fundamental Q-balls and their radial excitations}

If conditions (\ref{cond1})--(\ref{cond3}) are fulfilled, then the field
equation admits globally regular solutions $\phi(r)$ with finite energy
\cite{Coleman:1985ki}. The necessary condition for the energy
(\ref{radenergy}) to be finite is that the potential $U \to 0$ for
large $r$, and therefore $\phi \to 0$ as $r \to \infty$. Linearizing
Eq.~(\ref{radeom}) around $\phi = 0$, one finds that asymptotically
\begin{eqnarray}
\label{asymptrad}
\phi & = & \frac{A} {r} \exp \left\{-\sqrt{(U''(0) - \omega^2)} \, r
\right\} \left(1+O(1/r)\right) \;,
\end{eqnarray}
where $A$ is an integration constant. In view of (\ref{cond1}) the
argument of the exponent is real and negative, and so $\phi$ approaches
zero exponentially fast.

Solutions must also be regular at the origin of the coordinate system,
$r=0$. Since $r=0$ is the regular singular point of Eq.~(\ref{radeom}),
the solution will only be regular if it belongs to the `stable manifold'
characterized by the local Taylor expansion in the vicinity of $r=0$,
\begin{eqnarray}
\label{expanrad}
\phi(r) & = & \phi_0 + (U'(\phi_0)-\omega^2\phi_0) \, r^2 + O(r^4) \;,
\end{eqnarray}
where $\phi_0$ is an integration constant. 

Extending the two local solutions (\ref{asymptrad}),~(\ref{expanrad}) to
finite values of $r$ and requiring that $\phi$ and $\phi'$ for both
solutions agree at some $r = r_0$, yields two conditions for the free
parameters $A$ and $\phi_0$. Resolving these conditions determines a
globally regular solution in the interval $r \in [0,\infty)$. In fact,
in this way an infinite discrete family of globally regular solutions
parametrized by the number $n = 0,1,2,\ldots$ of nodes of $\phi$ is
obtained. The existence of these solutions can be illustrated by the
following qualitative argument. 

\begin{figure}[htb]
\begin{minipage}[t]{0.45\linewidth}
  \centering
  \includegraphics[width=\linewidth]{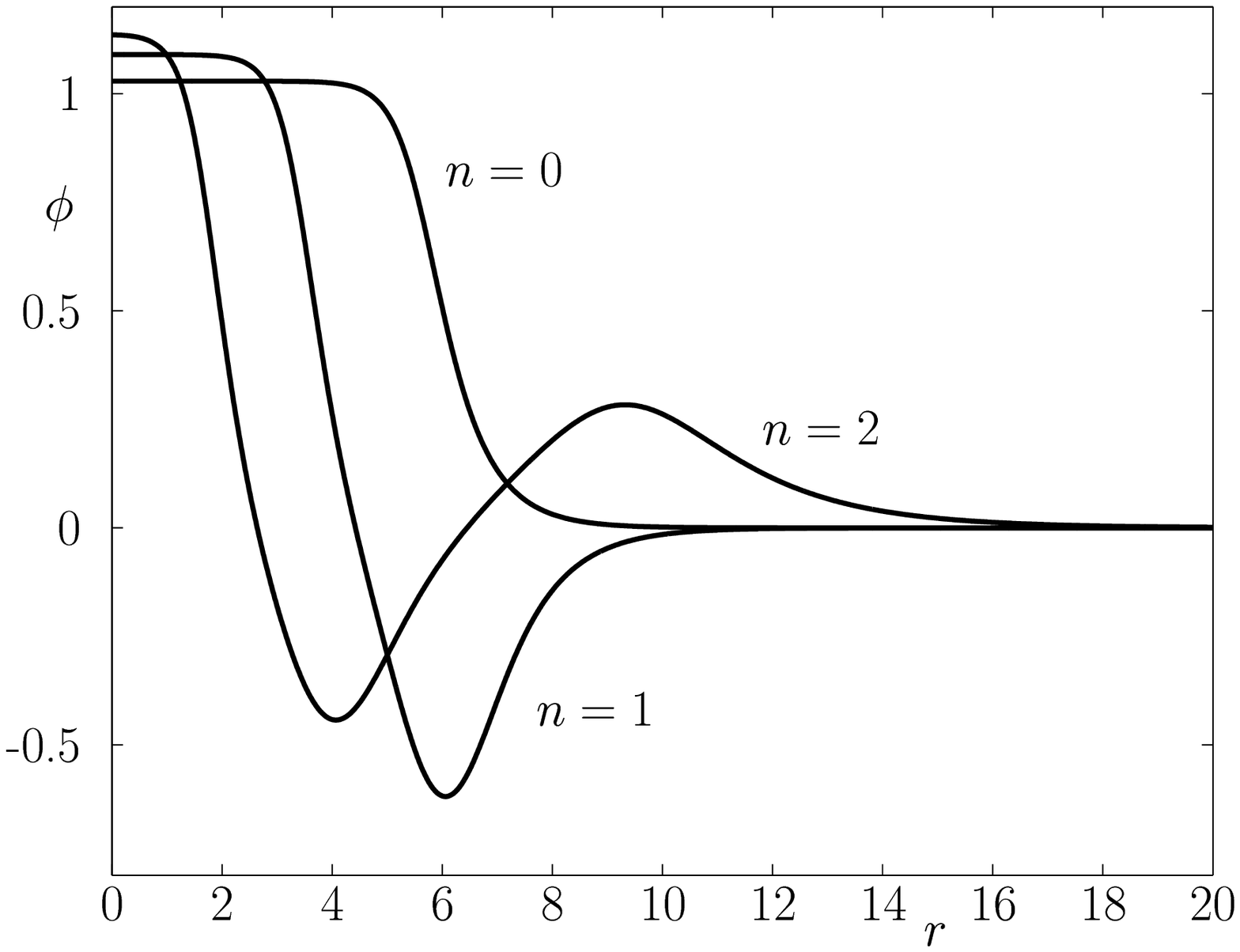}
  \caption{The profile of $\phi(r)$ for the fundamental Q-ball solution
  and its first two radial  excitations.}
  \label{fig3}
\end{minipage}
\hspace{5mm}
\begin{minipage}[t]{0.45\linewidth}
  \centering
  \includegraphics[width=\linewidth]{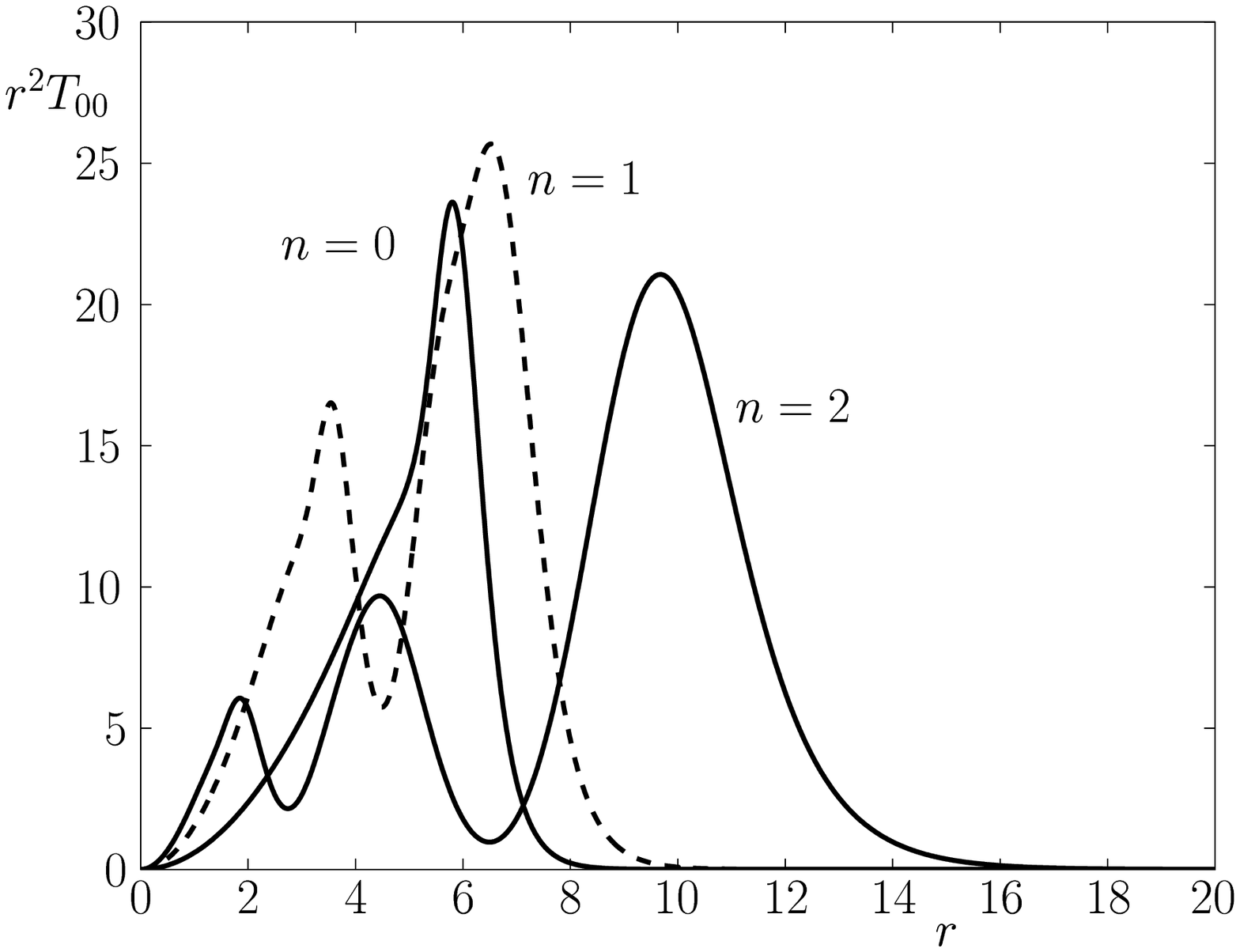}
  \caption{The radial energy density 
   $r^2T_{00}$ for the fundamental Q-ball and
  its first two excitations.}
  \label{fig4}
\end{minipage}
\end{figure}    

The parameter $\phi_0$ in (\ref{expanrad}) is the coordinate of the
`particle' at the initial moment of `time', $r=0$, when the particle
velocity is zero, $\phi'(0) = 0$. The particle therefore starts its
motion from some point on the curve $V(\phi)$ in Fig.~\ref{fig2}, with
the total effective energy $\CE$ being equal to its potential energy.
Then it moves to the right, dissipating some of its energy along its
way.
For $r \to \infty$, the particle must end up at the local maximum of the
potential $V$ (at $\phi = 0$) with 
the total effective energy being zero. This can be achieved by
fine-tuning the initial position of the particle, $\phi_0$. If
$V(\phi_0) < 0$, the effective 
energy of the particle will always be \textit{negative}, and therefore
it will not be able to end up in a configuration with zero energy. On
the other hand, if $\phi_0$ is such that the particle starts very close
to the absolute maximum of $V$, then it will stay there for a long
`time' $r$, during which period the dissipation term in (\ref{radeom2}),
which is $\sim 1/r$, will become very small. As a result, when the
particle will eventually start moving, its energy will be too large, and
so it will `overshoot' the position with zero energy. By continuity,
there is a value $\phi_0$ for which the total effective energy (the
right hand side in (\ref{radeom2})) is exactly zero for $r \to \infty$,
and so the particle will travel from $\phi(0) = \phi_0$ to $\phi(\infty)
= 0$ \cite{Coleman:1985ki}.

Next, one can fine-tune $\phi_0$ such that the initial energy
$V(\phi_0)$ is slightly too large, so that the particle first overshoots
the $\phi=0$ position, but then it hits the barrier from the other side,
bounces back and dissipates just enough energy to finally arrive at
$\phi=0$ with zero energy. This will give a solution with one node of
$\phi(r)$ in the interval $r \in [0,\infty)$. Similarly one can obtain
solutions with $n > 1$. To recapitulate, for each $\omega^2$ subject to
(\ref{cond1}), (\ref{cond2}) there is a solution to Eq.~(\ref{radeom})
for which $\phi(r)$ smoothly interpolates between some finite value at
the origin and zero value at infinity, crossing zero $n$ times
in between. We shall call solutions with $n = 0$ fundamental Q-balls.
Solutions with $n>0$ are the radial Q-ball excitations. The family of
Q-ball solutions can thus be parameterized by $(\omega,n)$.

As the frequency $\omega$ changes in the range $\omega_{\min}^2 <
\omega^2 < \omega_{\max}^2$, the charge
\begin{eqnarray}
\label{charge}
Q(\omega) & = & 8 \pi \omega \int_0^\infty \ud r \, r^2 \phi^2
\end{eqnarray}
changes from $Q(\omega_{\min}) = \pm \infty$ (depending on the sign of
$\omega$) to $Q(\omega_{\max}) = 0$. 
This can be understood as follows. For $\omega^2 \to \omega_{\max}^2$
the minimum of the effective potential $V(\phi)$ becomes more and more
shallow and moves closer and closer to $\phi = 0$. This implies that the
range of values of the solution $\phi(r)$ diminishes and the interval of
$r$ in which $\phi(r)$ is not constant decreases.  Q-balls therefore
`shrink' in this limit, and the integral (\ref{charge}) tends to zero%
\footnote{Although the exponent in (\ref{asymptrad}) decays slower and
slower for $\omega^2 \to U''(0)$, the prefactor $A$ vanishes faster
still.}. In the opposite limit, $\omega^2 \to \omega_{\min}^2$, Q-balls
become large, and their charge tends to infinity.

These considerations imply that instead of $\omega$ one can choose the
charge $Q$ as the independent parameter. Spherically symmetric $Q$-balls
therefore comprise a two-parameter family labeled by $(Q,n)$, where the
charge $-\infty < Q < \infty$ and the `excitation number' $n =
0,1,2,\ldots$. In Figs.~3--4 the profiles of the fundamental solution and
its first two radial excitations are shown for $Q = 1100$. As one can
see from the shape of the radial energy density, 
the $n$-th solution
has the structure of $n+1$ concentric spherical layers of energy%
\footnote{To our knowledge, solutions with $n > 0$ have not been
reported in the literature so far.}.

\section{Spinning Q-vortices.}

Our aim now is to show that  Q-balls admit spinning generalizations. For
this we modify 
the ansatz (\ref{radphi}) according to
\begin{eqnarray}
\label{phirot}
\Phi & = & \phi(r,\vartheta) \, e^{i \omega t + i N \varphi} \;,
\end{eqnarray}
where $N$ is an integer. 
This produces a non-zero component of the angular momentum along the
$z$-axis.
Inserting this ansatz into the field equation (\ref{eq}), the problem
reduces to a non-linear PDE for the function $\phi(r,\vartheta)$. Before
we start solving this equation, however, it is instructive to consider a
simpler system which effectively lives in $2+1$ dimensions. Then the
problem reduces to an ordinary differential equation (ODE).

Let us pass from spherical coordinates $(t,r,\vartheta,\varphi)$ to
polar coordinates $(t,\rho,z,\varphi)$, and then assume that the
field does not depend on $z$: 
\begin{eqnarray}
\label{phiaxial}
\Phi & = & \phi(\rho) \, e^{i \omega t + i N \varphi} \;. 
\end{eqnarray}
This will correspond to a vortex-type configuration, invariant under
translations along the $z$-axis. The energy per unit vortex length is
\begin{eqnarray}
\label{vortenergy}
E & = & 2 \pi \int_0^\infty \rho \, \left(\omega^2 \phi^2 + \phi^{\prime 2}
+ \frac{N^2} {\rho^2} \, \phi^2 + U(\phi)\right) \, \ud \rho \;.
\end{eqnarray}
If $N \neq 0$, the vortex will rotate around the $z$-axis, 
its angular momentum per unit length being given by
\begin{eqnarray}
\label{vortj}
J & = & \int T_{0\varphi} \, \rho \, \ud \rho \, \ud \varphi \; = \; 4
\pi \omega N \int_0^\infty \rho \, \phi^2 \, \ud \rho \; \equiv \; N Q
\;,
\end{eqnarray}
where $Q$ is the charge per unit length. The field equation now reads
\begin{eqnarray}
\label{vorteom}
0 & = & \phi'' + \frac{1} {\rho} \, \phi' - \frac{N^2} {\rho^2} \, \phi
- \frac{\ud U(\phi)} {\ud\phi} + \omega^2 \phi \;. 
\end{eqnarray}
For $N=0$ this reduces to (\ref{radeom}), with the replacement $2/r \to
1/\rho$ in the friction term. All arguments above still apply, hence we
conclude 
that there exist globally regular `Q-vortex'
solutions and their radial excitations with finite energy per unit length.   
These solutions display the behaviors qualitatively similar to those shown in 
Fig.~\ref{fig3}.

Let us now consider solutions with $N > 0$. As can be seen from
(\ref{vortenergy}), the energy for such solutions will be finite if only
$\phi(0) = 0$, such that the boundary condition for small $\rho$ is now
different. The power series solution to (\ref{vorteom}) in the vicinity of
$\rho=0$ reads
\begin{eqnarray}
\label{vortexpan}
\phi & = & B \rho^N + O(\rho^{N+1}) \;,
\end{eqnarray}
where $B$ is an integration constant. The asymptotic behavior for large
$\rho$ is
\begin{eqnarray}
\label{vortasympt}
\phi & = & \frac{A} {\sqrt{\rho}} \exp \left\{-\sqrt{(U''(0)-\omega^2)}
\, \rho \right\} \left(1 + O(1/\rho) \right) \;.
\end{eqnarray}
One can use essentially the same qualitative considerations as in the
preceding section to argue that globally regular solutions to
(\ref{vorteom}) with such boundary conditions exist, provided that
$\omega$ still fulfills conditions (\ref{cond1}), (\ref{cond2}). These
solutions can be obtained by numerically extending the asymptotics
(\ref{vortexpan}), (\ref{vortasympt}) to finite values of $\rho$ and
adjusting the free parameters $A$, $B$ to fulfill the matching
conditions at some $\rho=\rho_0$.

\begin{figure}[ht]
\begin{minipage}[b]{0.45\linewidth}
  \centering
  \includegraphics[width=\linewidth]{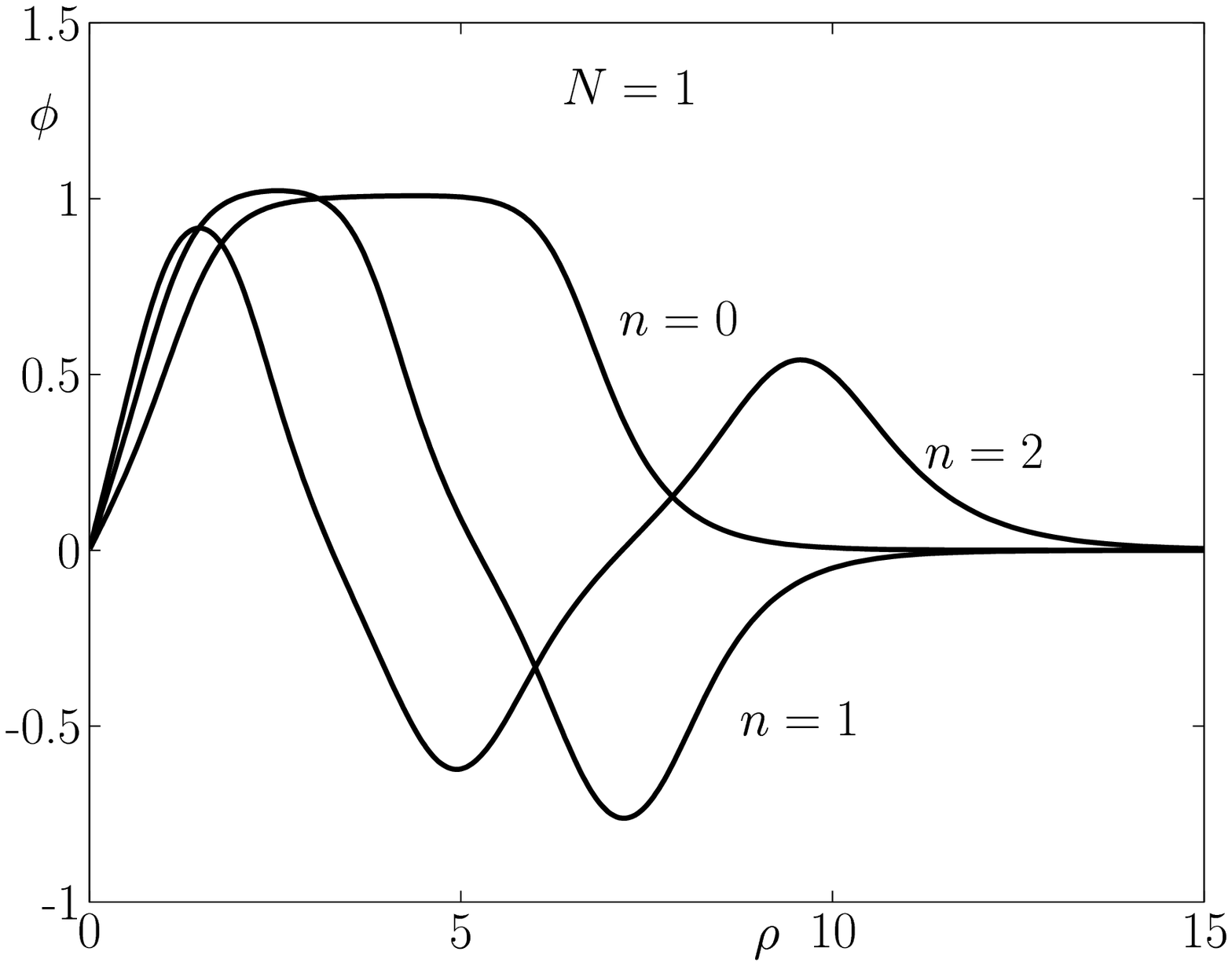}
  \caption{The spinning $N=1$ fundamental Q-vortex solution and its
  first two radial excitations; $Q=150$.}
  \label{fig5}
\end{minipage}
\hspace{5mm}
\begin{minipage}[b]{0.45\linewidth}
  \centering
  \includegraphics[width=\linewidth]{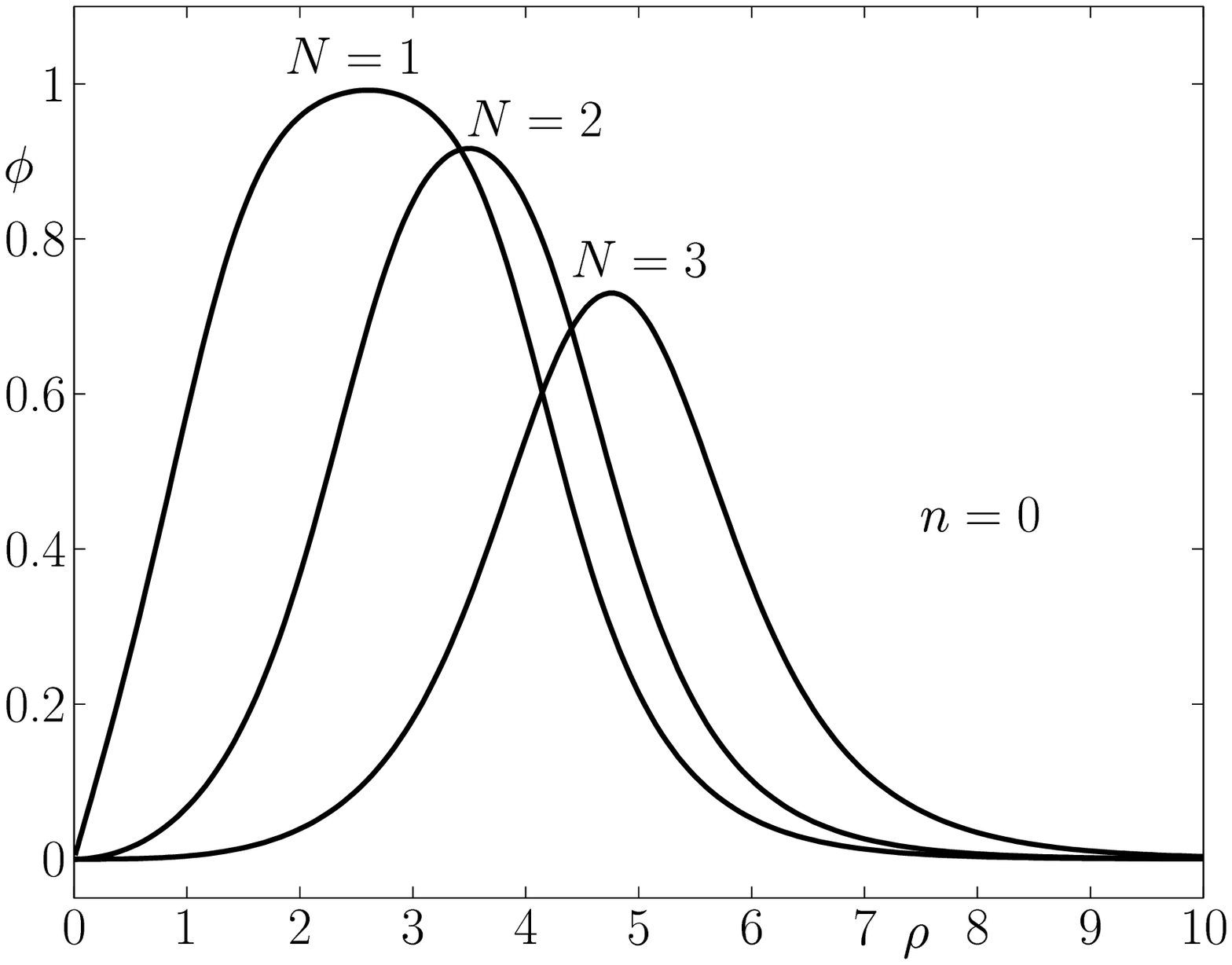}
  \caption{The first three spinning excitations of the fundamental
  Q-vortex solution; $Q=60$.}
  \label{fig6}
\end{minipage}
\end{figure}  

The conclusion is that there exists a family of globally regular
spinning Q-vortex solutions. These solutions can be labeled by
$(Q,n,N)$, where $Q$ is the charge per unit vortex length, $n =
0,1,2,\ldots$ is the radial `quantum' number (the number of nodes of
$\phi(\rho)$) and $N = 0,1,2,\ldots$ is the rotational `quantum'
number\footnote{Spinning Q-vortices with $n=0$ have also been found in 
Ref.~\cite{Kim:1993mm}.}.
For fixed $Q$, the energy per unit length, $E(Q,n,N)$, depends on
both $n$ and $N$, while the angular momentum is determined only by the
value of $N$ as $J = N Q$.  The profiles of $\phi(\rho)$ for several
low-lying excitations of the fundamental Q-vortex are shown in
Figs.~\ref{fig5}--\ref{fig6}.

\section{Spinning Q-balls}

Having considered the simpler problem in $2+1$ dimensions, we now return
to our main task of finding rotating Q-ball solutions in $3+1$
dimensions. Although, these two cases are qualitatively somewhat
similar, the $3+1$ dimensional problem is technically more
involved, since it requires solving a non-linear PDE. With $\Phi  =
\phi(r,\vartheta) \; e^{i\omega t+i N\varphi}$ the field equation
reduces to
\begin{eqnarray}
\label{eom3d}
\left( \frac{\partial^2} {\partial r^2} + \frac{2} {r} \frac{\partial}
{\partial r} + \frac{1} {r^2} \frac{\partial^2} {\partial \vartheta^2}
+ \frac{\cos\vartheta} {r^2 \sin\vartheta} \frac{\partial} {\partial
\vartheta} - \frac{N^2} {r^2 \sin^2\vartheta} + \omega^2 \right) \phi &
= & \frac{\ud U(\phi)}{\ud \phi} \;. 
\end{eqnarray}
The energy, $E = \int T_{00} \, \ud^3 x$, reads 
\begin{eqnarray}
\label{energy}
E & = & 2 \pi \int_0^\infty \ud r \, r^2 \, \int_0^\pi \ud
\vartheta \, \sin\vartheta \left(\omega^2 \phi^2 + (\partial_r \phi)^2 +
\frac{1} {r^2} (\partial_\vartheta \phi)^2 + \frac{N^2\phi^2} {r^2
\sin^2 \vartheta} + U(\phi) \right) \;, 
\end{eqnarray}
the charge
\begin{eqnarray}
\label{charge3d}
Q & = & 2 \omega \int \phi^2 \, \ud^3 x \; = \; 4 \pi \omega
\int_0^\infty \ud r \, r^2 \int_0^\pi \ud \vartheta \, \sin \vartheta \,
\phi^2 \;, 
\end{eqnarray}
and the angular momentum 
\begin{eqnarray}
\label{angmoment}
J & = & \int T_{0\varphi} \, \ud^3 x \; = \; N Q \;.
\end{eqnarray}
Finiteness of the energy requires that 
\begin{eqnarray}
\label{bound}
\phi & \to & 0 \qquad \text{as} \quad r \; \to \; 0 \text{ or } \infty
\;.
\end{eqnarray}
The asymptotic behavior of the solutions in these limits can be easily
determined, since for small $\phi$ one has $\ud U/\ud \phi \approx
U''(0) \, \phi$, such that equation (\ref{eom3d}) actually becomes
linear. The variables then separate, implying that the most general 
asymptotic solution has the form 
\begin{eqnarray}
\label{sep}
\phi(r,\vartheta) & = & \sum_{l=N}^\infty f_{l}(r) \,
P^N_l(\cos\vartheta) \;.
\end{eqnarray}
Here, $P^N_l(\cos\vartheta)$ are the associated Legendre functions. At
the origin, the radial amplitudes $f_l(r)$ are
\begin{eqnarray}
\label{expan}
f_l(r) & = & C_l \, r^l + O(r^{l+1}) \;,
\end{eqnarray}
while at infinity
\begin{eqnarray}
\label{asympt}
f_l(r) & = & \frac{A_l}{r} \, \exp \left\{-\sqrt{(U''(0)-\omega^2)} \, r
\right\} \left( 1 + O(1/r)\right) \;,
\end{eqnarray}
$C_l$ and $A_l$ denoting integration constants. 

Eqs.~(\ref{sep})--(\ref{asympt}) have been obtained by linearizing the
field equation in the vicinity of $r = 0$ and $r = \infty$, in which
case modes with different values of the quantum number $l$ decouple. Our
strategy to construct solutions in the whole space is to employ again 
the partial wave decomposition (\ref{sep}). This is always possible,
since the associated Legendre functions form a complete set.  However,
since for arbitrary values of $r$ the equation is non-linear, harmonics
with different values of $l$ will no longer decouple.

Since Eq.~(\ref{eom3d}) is symmetric with respect to reflections in
the $xy$-plane, $\vartheta \to \pi - \vartheta$, it follows that if
$\phi(r,\vartheta)$ is a solution, so is $\phi(r,\pi-\vartheta)$. In
addition, $-\phi(r,\vartheta)$ is also a solution, since the field
equation contains only odd powers of $\phi$. The associated Legendre
functions $P^N_l(\cos\vartheta)$ are even/odd with respect to 
$\vartheta \to \pi - \vartheta$ for even/odd values of $l+N$,
respectively. As a result, half of the terms in the mode expansion
(\ref{sep}) will change sign under the reflection, while the other half
will stay invariant. Since $\phi(r,\pi-\vartheta)$ must also be a
solution, it follows that either all odd or all even terms in the mode
expansion must vanish in order to have either $\phi(r,\pi-\vartheta) =
\phi(r,\vartheta)$ or $\phi(r,\pi-\vartheta) = -\phi(r,\vartheta)$. 
The conclusion is 
that for a given value of $N$ there are two solutions with
different parities $P$:
\begin{eqnarray}
\label{sepeven}
P \; = \; +1: \quad \phi(r,\vartheta) & = & \sum_{k=0}^\infty f_{k}(r)
\, P^N_{N+2k}(\cos\vartheta) \;, \\
\label{sepodd}
P \; = \; -1: \quad \phi(r,\vartheta) & = & \sum_{k=0}^\infty f_{k}(r)
\, P^N_{N+2k+1}(\cos\vartheta) \;.
\end{eqnarray}

With our choice of the potential (\ref{polynpot}), the field equation
(\ref{eom3d}) contains cubic and quintic non-linearities. In view of
the completeness of the associated Legendre functions, their products
can be expressed in terms of their
linear combinations, for example, $\left(\sum f_l P_l^N(x) \right)^5 =
\sum a_j P_j^N(x)$, with the coefficients $a_j$ determined by the
$f_l$'s.
As a result, inserting (\ref{sepeven}),~(\ref{sepodd}) into
(\ref{eom3d}) we obtain
\begin{eqnarray}
\label{diffform}
\sum_{k}^\infty \CD_k[f_{s}(r)] \, P^N_{N+k}(\cos\vartheta) & = & 0 \;. 
\end{eqnarray}
Here $\CD_k[f_{s}(r)]$ are second order differential operators acting on
the radial amplitudes $f_s(r)$, and the sum is taken over all odd/even
positive $k$ for odd/even solutions, respectively. This equation is
equivalent to the infinite set of ODEs
\begin{eqnarray}
\label{diffset}
\CD_k[f_{s}(r)] & = & 0 \;. 
\end{eqnarray}
Now, we 
truncate this system by setting all amplitudes $f_s$ with $s$ larger
than some maximal value $l_{\max}-N$ to zero and by discarding all
equations with $k > l_{\max}-N$. The indices in (\ref{diffset}) then
vary only in the finite limits
\begin{eqnarray}
k,s & = & 0,1,\ldots, l_{\max}-N \;.
\end{eqnarray}
As a result, we end up with a finite system of ODEs. This procedure
is sometimes called Galerkin's projection method. It is natural to
expect that if $l_{\max}$ is large enough, the resulting approximate
solutions will be close to the exact ones. To illustrate that this is
indeed the case, we show in Table I the energy and charge of
the solution of the truncated system with $\omega^2 = N = P = 1$
for several values of $l_{\max}$. 
\begin{table}[htb]
\caption{Parameters of the solitons versus the truncation parameter 
$l_{\max}$.}
\label{tab1}
\centering
\vglue 0.4 cm
\begin{tabular}{|c|*{5}{c|}}
\hline
$l_{\max}$ & $1$    & $3$    & $5$    & $7$    & $9$    \\ 
\hline
$E$        & 230.70 & 217.16 & 207.74 & 205.58 & 205.36 \\
$Q$        & 211.46 & 200.68 & 189.43 & 186.99 & 186.73 \\
\hline
\end{tabular}
\end{table}
As one can see, the energy and charge indeed approach some limiting
values with growing $l_{\max}$. It actually seems to be sufficient
to take into account only the first 3--5 lowest harmonics in order to
get a reasonable approximation.

\begin{figure}[ht]
\begin{minipage}[t]{0.45\linewidth}
\centering
\includegraphics[width=\linewidth]{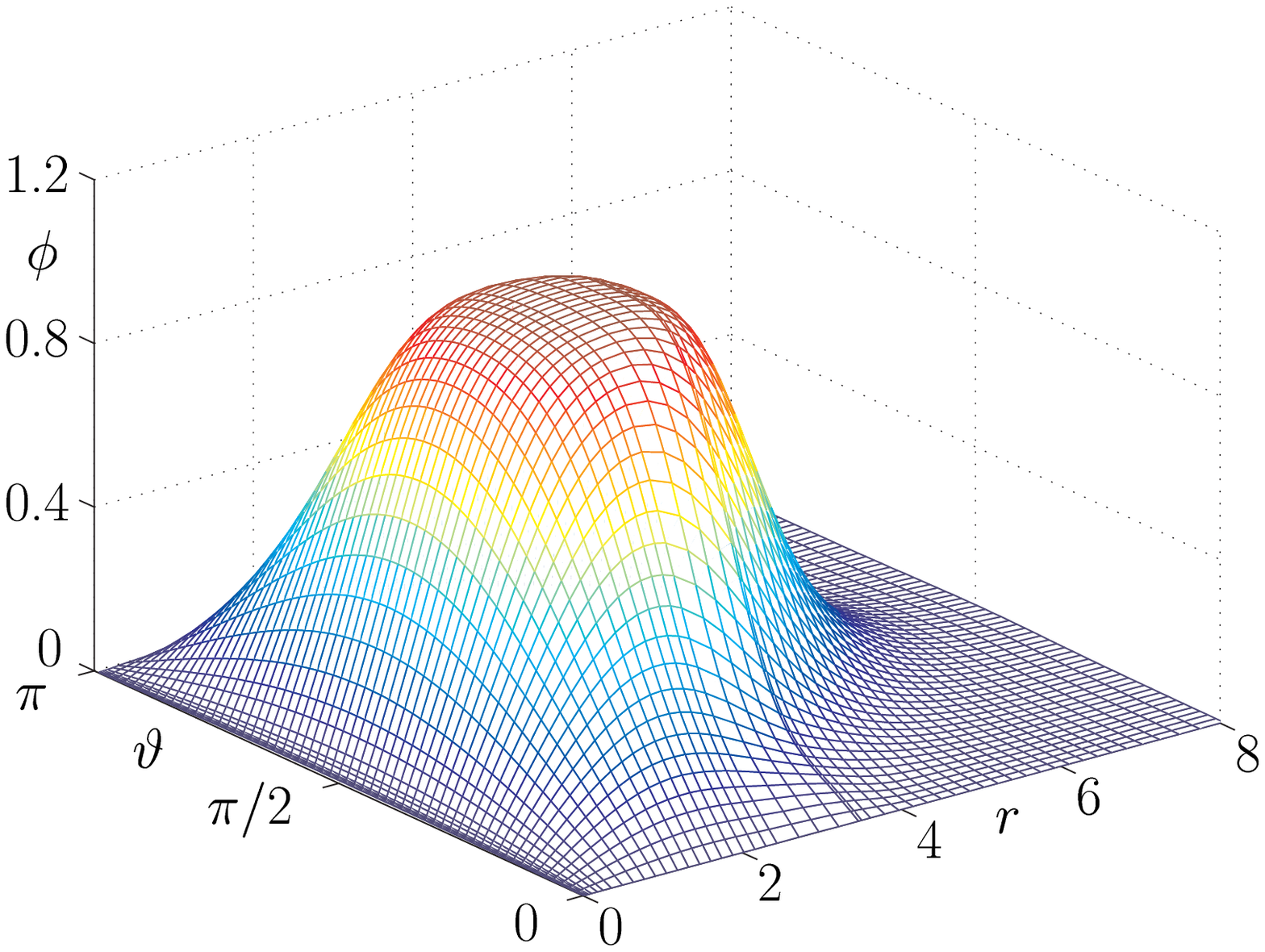}
\caption{$\phi(r,\vartheta)$ for $N = 1$, $P = 1$.}
\label{phi1}
\end{minipage}
\hspace{5mm}
\begin{minipage}[t]{0.45\linewidth}
\centering
\includegraphics[width=\linewidth]{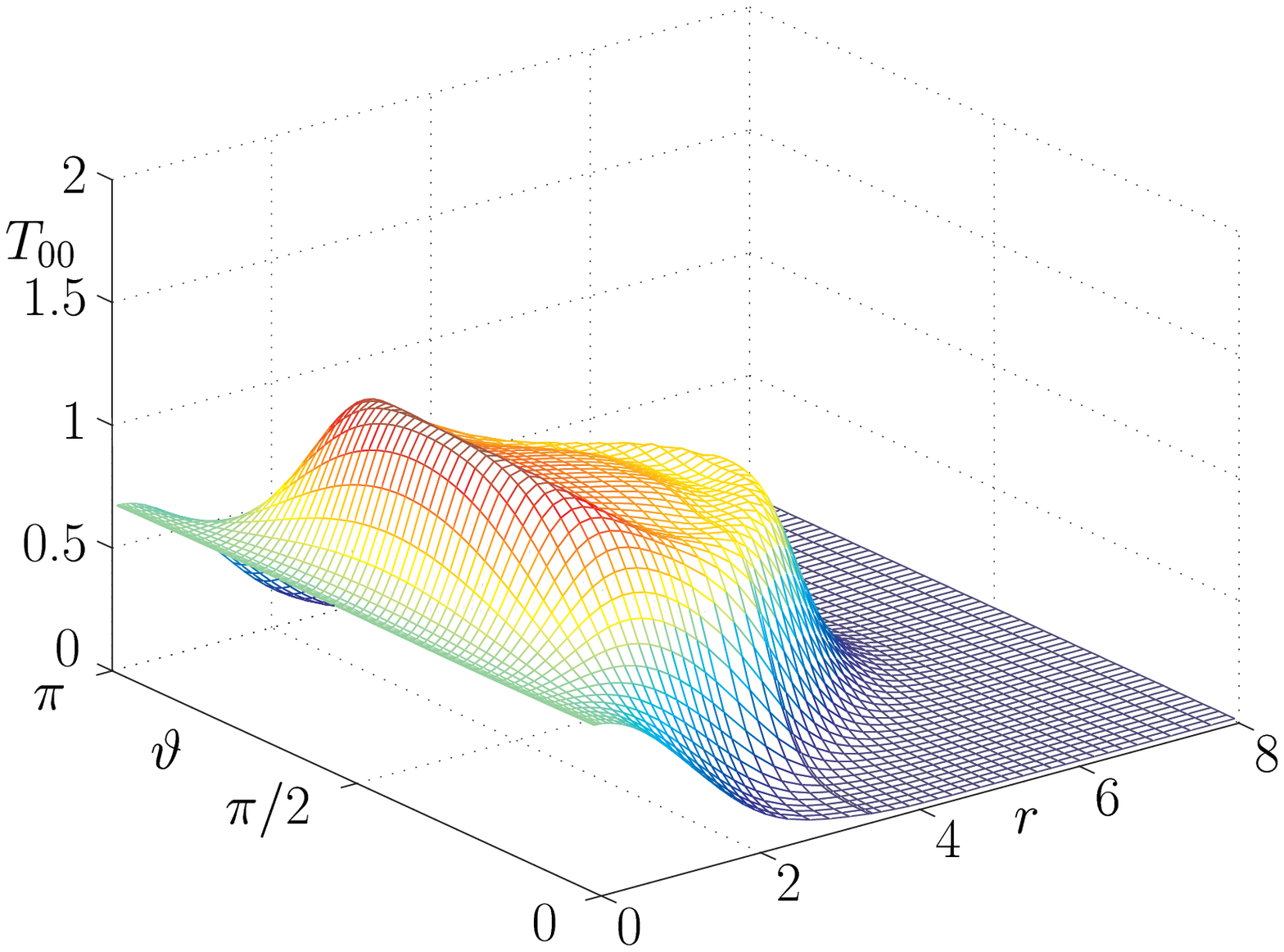}
\caption{$T_{00}$ for $N = 1$, $P = 1$.}
\label{t1}
\end{minipage}
\end{figure}

\begin{figure}[ht]
\begin{minipage}[t]{0.45\linewidth}
\centering
\includegraphics[width=\linewidth]{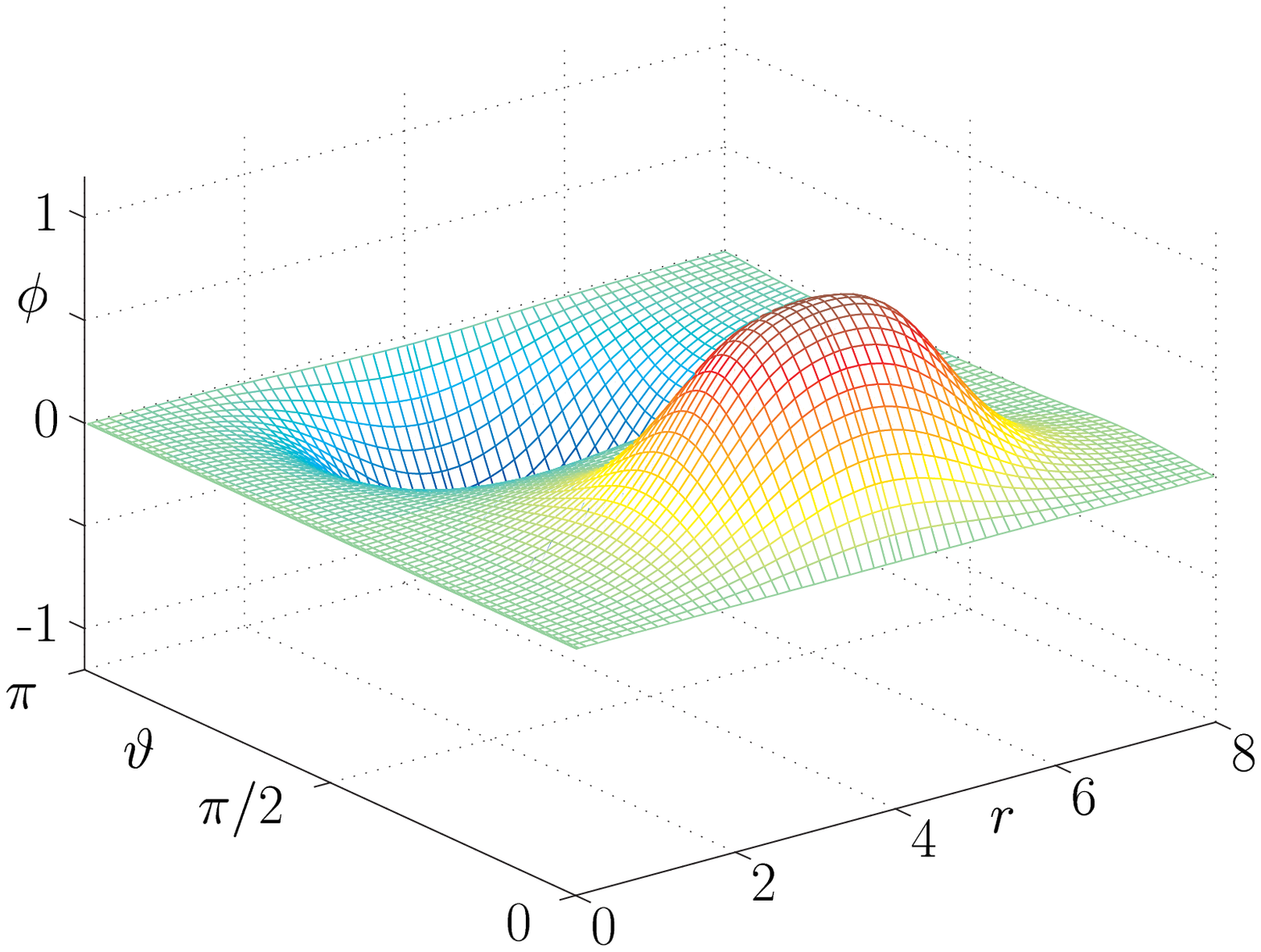}
\caption{$\phi(r,\vartheta)$ for the $N = 1$, $P = -1$.}
\label{phi1odd}
\end{minipage}
\hspace{5mm}
\begin{minipage}[t]{0.45\linewidth}
\centering
\includegraphics[width=\linewidth]{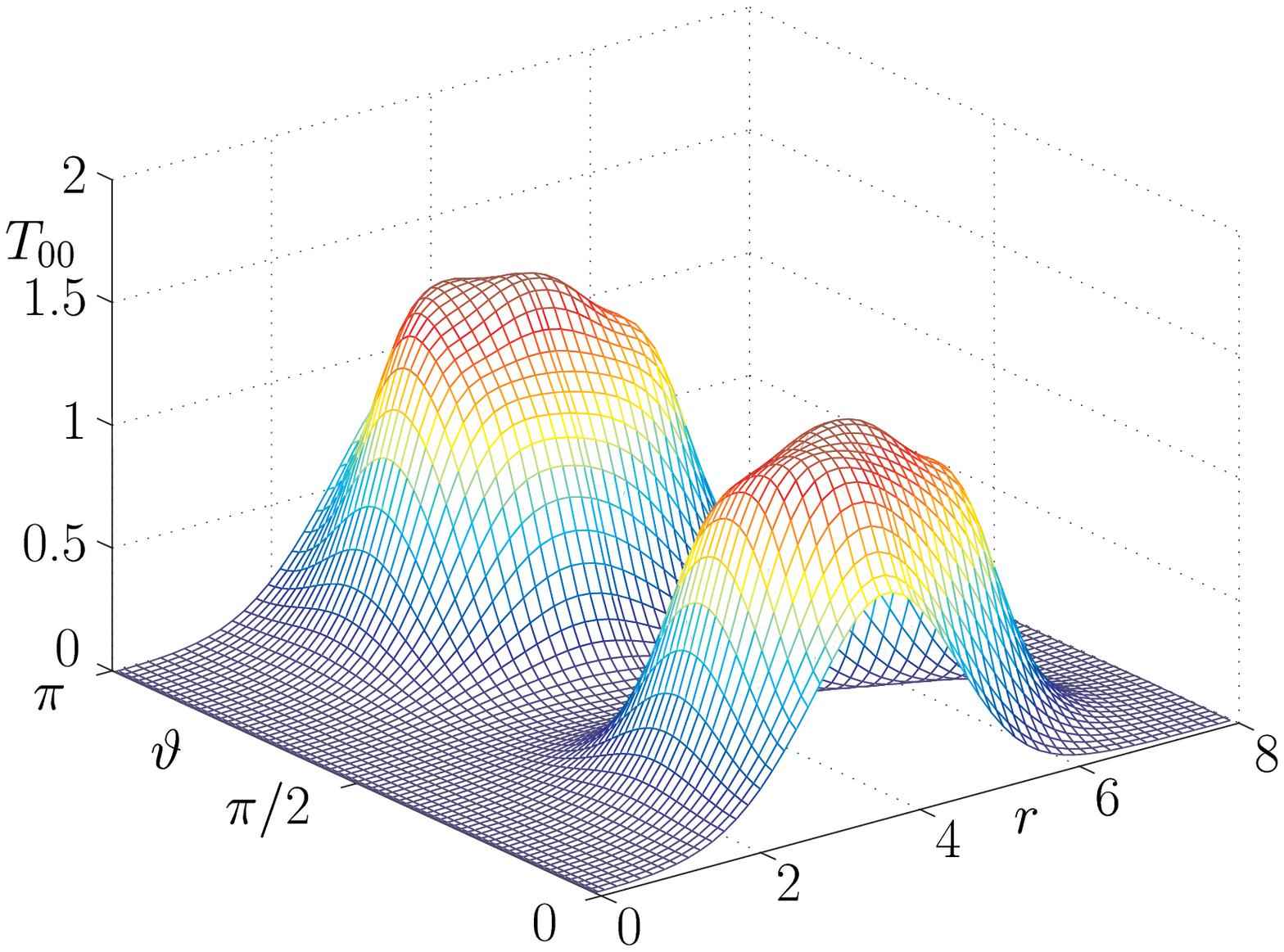}
\caption{$T_{00}$ for the $N = 1$, $P = -1$. }
\label{t1odd}
\end{minipage}
\end{figure}

In the next step we construct solutions for a fixed charge $Q$ in
different $N$ sectors. More precisely, the numerical solutions to
Eqs.~(\ref{diffset}) have been obtained with \matlab's ODE solvers
by utilizing the shooting method. The asymptotic solutions
(\ref{expan}),~(\ref{asympt}) were used to start the integration at
$r=0.01$ and at $r=10$ toward the matching point, whose position has
been varied between $r=3$ and $r=6$. The matching conditions imposed at
this point determine the values of the constants $C_l$ and $A_l$ in
(\ref{expan}) and (\ref{asympt}). The typical matching error was found
to be less then $10^{-16}$. The profiles of even and odd rotating
solutions with $N = 1$ are shown in Figs.~\ref{phi1}--\ref{t1odd}.
As one can see from these plots, the distribution of the energy density
$T_{00}$ is strongly non-spherical. It has the structure of a deformed
ellipsoid for $P = 1$, and that of dumbbells oriented along the
rotation axis for $P = -1$; in the latter case the energy density vanishes
in the equatorial plane. The energies of the first three rotational
excitations of the fundamental Q-ball are given in Table~\ref{tsq}. 
\begin{table}[htb]
\caption{Parameters of the rotating solutions with $Q = 410$.}
\label{tsq}
\centering
\vglue 0.4 cm
\begin{tabular}{|c|c|c|c|}
\hline
$N^{P}$   & $l_{\max}$ & $E$    & $\omega$ \\
\hline
$0$       &       0    & 307.29 &  0.76099 \\
\hline
$1^{+}$   &       9    & 378.36 &  0.86017 \\
\hline
$1^{-}$   &       8    & 442.24 &  0.99164 \\
\hline
$2^{+}$   &      10    & 433.94 &  0.96258 \\
\hline
$2^{-}$   &       9    & 505.66 &  1.13284 \\
\hline
$3^{+}$   &       9    & 473.59 &  1.05277 \\
\hline
$3^{-}$   &      12    & 528.30 &  1.31037 \\
\hline
\end{tabular}
\end{table}
As one can see, the energy of the first excitation exceeds the ground
state energy by about $20 \%$, the next excitation lying again about $20
\%$ above. 
This is in agreement with the expected properties of spinning
excitations of a single soliton. In summary, spinning Q-balls comprise a
two-parameter family labeled by the values of $(Q,N)$. We have also
found evidence for the existence of spinning radial excitations with $n
> 0$.  However, the construction of such solutions is somewhat more
involved and we refrain from presenting them in this paper.

\vspace{-\baselineskip}

\section{Conclusions}

The aim of this paper was to find an example of spinning solitons in
Minkowski space. We have considered the model of a complex scalar field
with a non-renormalizable self-interaction. In the spherically symmetric
sector this model contains non-topological solitons, the Q-balls. In
addition, we have found an infinite discrete family of radial Q-ball
excitations, parameterized by the number of nodes $n$ of the radial field
amplitude. Such excited solutions have not been reported in the
literature before.

In a second step we have analyzed cylindrically symmetric solutions with
explicit harmonic dependence on the azimuthal angle, $\exp(i N
\varphi)$, which we call spinning Q-vortices. For such solutions there
is a non-zero component of the angular momentum along the $z$-axis, $J =
N Q$, where $Q$ is the charge per unit vortex length. In addition, these
solutions exhibit radial excitations parameterized by an integer $n$. As
a result, such spinning solutions comprise a three-parameter family
labeled by $(Q,N,n)$.

Finally, we have considered the full $3+1$ dimensional problem. We have
used a version of the spectral method by expanding the field with
respect to the complete set of associated Legendre functions. We reduced
the PDE to an infinite system of radial ODEs, and then truncating this
system at finite order. 
The parameters of the solutions for the truncated system converge
rapidly to some limiting values as the truncation parameter grows. By
keeping the charge $Q$ fixed, we have obtained the lowest rotational
excitations of the fundamental Q-ball solutions. The angular momentum of
these solutions is quantized, $J = N Q$, the energy increases (but not
very rapidly) with the angular momentum. As a result, these solutions
can be viewed as describing \textit{spinning} excitations in the
one-soliton sector rather than orbital motion in a many-soliton system%
\footnote{One could imagine a situation where in the one-soliton sector
there is a soliton plus an orbiting soliton-antisoliton pair.  However,
the spectrum of $J$ would then probably be continuous, while the energy
would probably be considerably higher than that for the single
soliton.}. To our knowledge, these spinning Q-balls provide the
first explicit example of spinning solitons in Minkowski space in $3+1$
dimensions.

\begin{acknowledgments}

M.S.V. would like to acknowledge useful discussions with Dieter Maison
during the early stages of this research,
to thank Peter Forgacs for some interesting remarks, 
 and to thank Fidel Schaposnik for
bringing Refs.~\cite{deVega:1986eu,Jackiw:1990aw} to our attention.  We
would also like to thank Andreas Wipf for his help and useful comments,
and also Tom Heinzl for a careful reading of the manuscript. 
The work of E.W. was supported by the DFG grant Wi 777/4-3. The work of
M.S.V. was supported by CNRS.

\end{acknowledgments}

\end{document}